\documentstyle[12pt,epsfig]{article}

\begin{document}

  \newcommand{\ccaption}[2]{
    \begin{center}
    \parbox{0.85\textwidth}{
      \caption[#1]{\small{{#2}}}
      }
    \end{center}
    }

\hfill {\sf IHEP--97--79} 

\vspace{0.5cm}

\begin{center}
{\large 
{\bf {\Large { \sf Search for Anomalous Top--Quark Interaction at LEP--2 
Collider }}}}
\end{center}

\vspace{1cm}

\begin{center}
V.F.~Obraztsov$^1$, S.R.~Slabospitsky$^2$, and O.P.~Yushchenko$^3$
\end{center}

\vspace{1cm}

\begin{center}

State Research Center \\
Institute for High Energy Physics, \\
Protvino, Moscow Region 142284 \\
Russia

\end{center}

\vspace{1cm}

We show that a search for $e^+ e^- \to t \bar q$ $(\bar q =
\bar c, \bar u)$ events at LEP--2 collider provide a possibility
to improve significantly  the modern constraints on coupling 
constants of anomalous $t$--quark interaction
via flavor--changing neutral currents.

\rule{3cm}{0.5pt}

$^1$E--mail:~OBRAZTSOV$@$mx.ihep.su,

$^2$E--mail:~SLABOSPITSKY$@$mx.ihep.su

$^3$E--mail:~YUSHCHENKO@mx.ihep.su
\newpage

\section{\bf Introduction }

The discovery of the $t$-quark on FNAL collider~\cite{disc} opens new
experimental direction in search for new physics beyond the Standard 
Model~(SM). 

A search for rare decays of a top quark is one of such studies. Of special
interest is the search for $t$ quark decays via flavor--changing neutral 
neutral currents (FCNC--decays)~\cite{parke}:
\begin{eqnarray}
 t \, \, \to \, \, \gamma \, (g, \, Z) \; \; + \; \; c \, (u). \label{1} 
\end{eqnarray}

At the tree level of SM there are no vertices, corresponding to these
FCNC--processes. Only "loop" contributions make possible the 
decays~(\ref{1}). As
a result the branching fractions of those decays are very small~\cite{sm}:
\begin{eqnarray}
 {\rm Br} ( t \to (\gamma, \, g, \, Z) \; + \; c(u) ) < 10^{-10}.  \label{2} 
\end{eqnarray}

However, many extensions of SM lead to huge enhancement of such transitions.
Therefore, the observation of FCNC--decays of the $t$--quark would be an
evident indication of new physics beyond the SM (see, for 
example,~\cite{parke,th1,th2,arbuzov}).

CDF collaboration perform the search for decays (\ref{1}) at FNAL collider in
$\bar p p$ collisions at the energy of $\sqrt{s} = 1.8$~TeV in the reaction of
top quarks production:
\begin{eqnarray}
 \bar p p \; \to \; \bar t t X. \label{recpp}
\end{eqnarray}
The collaboration obtain the upper limits on branching fractions for the
decays $t \to \gamma c(u)$ and $t \to Z c(u)$~\cite{cdf} as follows:
\begin{eqnarray}
 {\rm Br}(t \to c \gamma) + {\rm Br}(t \to u \gamma) < 3.2\% 
\quad (95\% \;\;{\rm CL}), \label{dec1} \\
 {\rm Br}(t \to c Z) + {\rm Br}(t \to u Z) < 33\% 
\quad (95\% \;\;{\rm CL}). \label{dec2}
\end{eqnarray}

These "feeble" constraints (that is approximately $\sim 10^8$ tines more than
SM predictions~\cite{sm}) are resulted naturally due to small collected
statistics of events with $t$ quark production ($N_{\bar t t} \sim 10^2$). In
future run of the FNAL collider it is expected a significant increase of
statistics ($N_{\bar t t} \sim 10^3 \div 10^4$~\cite{fut}), which would make
possible to improve the estimate~(\ref{dec1}) and~(\ref{dec2}) 
(see~\cite{th2,fut} for details). 

In this article we consider a possibility to evaluate the similar constraints
on coupling constants of anomalous $t$ quark interaction from the data of
$e^+ e^-$ collider LEP-2 at CERN. We study the process of a single top quark
production via FCNC interactions:
\begin{eqnarray}
e^+ \; e^- \; \to \; \gamma^{\ast} (Z^{\ast}) \; \to \; t \;\; \bar c (\bar u).
\label{etc1} 
\end{eqnarray}

Note, that this process~(\ref{etc1}) was considered early (see, for 
example,~\cite{eetc} and references therein). However, these articles studied
the manifestation of such an interaction at the energies of future $e^+ e^-$
collider ($\sqrt{s} \sim 500~$çÜ÷), while a detail investigation of the
reaction~(\ref{etc1}) at the energies of LEP-2 collider was not performed.

Beginning in summer 1997, $e^+ e^-$ collider LEP-2 operates at the energy of
$\sqrt{s}$ = 184~çÜ÷. At such a total energy of $e^+ e^-$ annihilation it is
kinematically possible the production of single top quarks in the 
reaction~(\ref{etc1}). Thus a search for such process becomes a quite
reasonable problem.

In this article we study the problem as follows: what are the constraints
on anomalous coupling constants of FCNC interaction of the top quark can be 
extracted from the data of collider LEP-2 ? We exhibit that at the projected
luminosity of ${\cal L}_{e^+ e^-} \sim 100$~pb${}^{-1}$ one can improve
significantly the current constraints on the magnitudes of anomalous coupling
constants, which following from~(\ref{dec1}) and~(\ref{dec2}). 

The article is organized as follows. The anomalous FCNC interaction of $t$
quark is considered in Section~2. The estimates on $t$ quark yields and the
constraints on anomalous couplings are presented in Section~3. The
differential distributions with respect to energies and the outgoing angles of
final particles are given in Section~4. The main
results are summarized in Conclusion.

\section {\bf $\gamma \bar t c$ and $Z \bar t c$ vertices }

Let us present the form of the anomalous vertices of flavor-changing neutral
currents. We consider of $V_0 \bar t c$ and $V_0 \bar t u$ vertices, where 
$V_0$ stands for a photon or $Z$--boson. For definiteness sake, we describe 
the $t$ quark transition into charmed $c$ quark. Top transition into $u$ quark
is described in a similar way.

Following~\cite{th1}, the vertices of the FCNC transitions $\gamma \to f 
\bar{f'}$ and $Z \to f \bar{f'}$ can be written as follows:

\begin{eqnarray}
\Gamma_{\mu}^{\gamma} & = & \kappa_{\gamma} \frac{ee_q}{\Lambda} 
\sigma_{\mu\nu} \left( g_1 P_l + g_2 P_r \right ) q^{\nu}, \label{ver1} \\
\Gamma_{\mu}^{Z} & = & \kappa_{Z} \frac{e}{\sin 2\vartheta_W} 
\gamma_{\mu} \left( z_1 P_l + z_2 P_r \right ) \label{ver2} 
\end{eqnarray}
where $\Lambda$ is the new physics cutoff; $e$ is the electric
charge, $e_q = 2/3$ is the charge of $t$ quark, $\vartheta$ is the Weinberg
angle, $\sigma^{\mu \nu} = \frac{1}{2}(\gamma^{\mu} \gamma^{\nu} - 
\gamma^{\nu} \gamma^{\mu})$, 
$P_{\frac{l}{r}} = \frac{1}{2}(1 \pm \gamma^5)$, 
$\kappa_{\gamma}$ and $\kappa_z$ define the
strength of the anomalous couplings for the current with a photon 
($\kappa_{\gamma}$)
and $Z$ boson ($\kappa_z$), respectively. The relative magnitudes of right and
left components of the currents are denoted by $g_1$, $g_2$, $z_1$, 
and $z_2$. They obey the obvious constraints as follows:
\begin{eqnarray}
 g_1^2 + g_2^2 = 1, \quad  z_1^2 + z_2^2 = 1. \label{rel} 
\end{eqnarray}
We also assume that ${\rm Im} \kappa_{\gamma} = {\rm Im} \kappa_z 
= {\rm Im} g_i = {\rm Im} z_i = 0$, see~\cite{th1,th2}.

Since all equations in our article contain the mass scale parameter
$\Lambda$ in combination of $\kappa_{\gamma} / \Lambda$, then for definiteness
sake in what follows we assume that 
\[ 
\Lambda = m_t. \label{ver3} 
\]
Using the expressions for the vertices~(\ref{ver1}) and (\ref{ver2}) for the
corresponding widths we find (see, also~\cite{th2}): 
\begin{eqnarray}
\Gamma( t \to c \gamma) &=& \kappa^2_{\gamma} \frac{\alpha e_q^2}{4} \Biggl (
 \frac{m^2_t}{\Lambda^2} \Biggr ) m_t, \label{ver4} \\
\Gamma( t \to c Z) &=& \kappa^2_{z} 
\frac{\alpha}{8 \sin^2 2\vartheta_W \, M^2_Z} 
m_t^3 \Biggl ( 1 - \frac{M_Z^2}{m_t^2} \Biggr )^2 
\Biggl ( 1 + 2\frac{M_Z^2}{m_t^2} \Biggr ),
\label{ver5}
\end{eqnarray}
where  $\alpha$ is the fine structure constant, $M_Z$ is the mass of $Z$
boson.

We put the mass of the light quark ($c$ or $u$) equals zero, $m_c = m_u = 0$, 
in the equations~(\ref{ver4}) and~(\ref{ver5}), since $m_q \ll m_t$. In all
estimates we use also
\begin{eqnarray}
m_t = 175 \, \; {\rm GeV}. \label{ver6}
\end{eqnarray}
That is in agreement with the latest experimental data~\cite{mt}:
\begin{eqnarray*}
 D\emptyset \quad m_t &=& 173.3 \pm 5.6(stat.) \pm 6.2(syst.) \;\; 
{\rm GeV}/c^2, \\
 CDF \quad m_t &=& 175.9 \pm 4.8(stat.) \pm 4.9(syst.)  \;\; {\rm GeV}/c^2.
\end{eqnarray*}

Using the equations~(\ref{ver4}) and (\ref{ver5}) from the experimental
constraints~(\ref{dec1}) and (\ref{dec2}) it is easily to obtain the 
magnitudes of upper limits on the constants of $\kappa_{\gamma}$ and $\kappa_z$ 
(at $m_t$ = 175~GeV): 
\begin{eqnarray}
 \kappa_{\gamma}^2 &<& 0.176 \quad {\rm at} \quad \Lambda = m_t, 
\label{const1} \\
 \kappa^2_z &<& 0.533. \label{const2} 
\end{eqnarray}

\section {\bf Total cross section for $t \bar q$ pair production in $e^+ e^-$
annihilation and constraints on magnitudes of anomalous coupling constants }

By means of the anomalous vertices~(\ref{ver1}) and (\ref{ver2}) one can
readily find the expression for the total cross section for $t$ and 
$\bar c (\bar u)$ quarks production in the reaction~(\ref{etc1}) (with
$m_c = 0$): 
\begin{eqnarray}
&&\sigma(e^+ e^- \to t \bar c)  =  \frac{\pi \alpha^2}{s} 
\Biggl ( 1 - \frac{m^2_t}{s} \Biggr )^2 
\Biggl [ \kappa^2_{\gamma} \frac{m^2_t}{\Lambda^2} e^2_q \frac{s}{m^2_t}
 \biggl ( 1 + \frac{2m^2_t}{s}\biggr )  \nonumber \\
 &+& \frac{ \kappa^2_{z} (1+a^2_w)(2+\frac{m^2_t}{s})}
{4 \sin^4 2\vartheta_W (1-\frac{M^2_Z}{s})^2} 
 + 3 \kappa_{\gamma} \kappa_z \biggl ( \frac{m_t}{\Lambda} \biggr ) 
\frac{ a_w e_q (g_1 z_1 + g_2 z_2)}
{\sin^2 2\vartheta_W (1-\frac{M^2_Z}{s})} \Biggr ], \label{sig0} 
\end{eqnarray}
where 
$a_w = 1 - 4 \sin^2 \vartheta_W$, the rest parameters are previously
described.

In this equation the first and second terms correspond to annihilation via
photon $(\sim \kappa^2_{\gamma})$ and $Z$ boson $(\sim \kappa^2_z)$,
respectively, while the third term $(\sim \kappa_{\gamma} \kappa_z)$ describes
their interference.

As one would expect, the behavior of a single top production cross section has
an evidently threshold character (see~(\ref{sig0})):
\[ 
\sigma(e^+ e^- \to t \bar c) \propto \biggl (1 - \frac{m^2_t}{s} \biggr )^2.
\]
Therefore, when evaluating the cross section for the process~(\ref{etc1}) at
the near-threshold region, one should take into account the finite widths of
$t$ quark and $W$ boson. In other words, at the energy of 
$\sqrt{s} \simeq m_t$ one should consider a virtual $t^{\ast}$ quark
production with a subsequent decays into a virtual $W^{\ast}$ boson:
\begin{eqnarray}
e^+ \; e^- \; \to \; \bar c (\bar u) \; t^{\ast} \;(\to b \; 
W^{\ast} \;(\to l \nu (q \bar q')). \label{etc2} 
\end{eqnarray}
The expression for matrix element of such process is too cumbersome and we do
not present it.

Fig.~1 exhibits the behavior of the cross sections for the
process~(\ref{etc1}) and (\ref{etc2}) as a function of~$\sqrt{s}$. As is seen
from this figure the effect of finite widths of $t$ quark and $W$ boson
manifests itself at the region of $\sqrt{s} \leq m_t$. At higher energies of
$e^+ e^-$ annihilation these two cross sections becomes practically equal in
magnitude. Because we examine the process~(\ref{etc1}) at the energy 
of~$\sqrt{s} \geq 184$~çÜ÷, then the basic characteristics of the reaction of
a single top production via FCNC interaction can be understood from analysis
of the equation~(\ref{sig0}) for the cross section for the 
process~(\ref{etc1}).  

Fig.~2 presents the behavior of the cross section production for the process
$e^+ e^- \; \to t \bar c$ as a function of~$\sqrt{s}$. We also show the
individual contributions corresponding to the exchange via virtual
$Z$ boson, virtual photon, and their interference. The presented estimates
of the cross sections are evaluated by use of the anomalous constants 
($\kappa_{\gamma}$ and $\kappa_z$) equal their "upper" values
(see~(\ref{const1}) and (\ref{const2})). As is seen this choice of the
constants magnitudes results in the dominating contribution of virtual $Z$
boson into cross section of the process~(\ref{etc1}) at 
$\sqrt{s} \leq 400$~GeV. We point out the difference in the energy behavior of
the contributions due to exchange via photon and $Z$ boson. Because of
anomalous magnetic interaction with a photon
($\sim \sigma^{\mu \nu}$), its contribution is not decrease with increasing of
the total energy of interaction. Indeed, from~(\ref{sig0}) one find 
\begin{eqnarray*}
\sigma(e^+ e^- \to \gamma^{\ast}) & \propto &
\biggl (1 - \frac{m^2_t}{s} \biggr) ^2, \\
\sigma(e^+ e^- \to Z^{\ast}) & \propto &
\frac{1}{s} \biggl (1 - \frac{m^2_t}{s} \biggr) ^2.
\end{eqnarray*}

At the energy of $\sqrt{s}$~=~184 GeV, that is corresponding to operation
energy of 1997 run of LEP-2 collider, the magnitude of the cross section
for the reaction~(\ref{etc1}) (summed over $t$ and $\bar t$ as well as over
$c$ É $u$ quarks) is equal to:
\begin{eqnarray}
\sigma(e^+ e^- \to t \bar c \;+\; t \bar u  \;+\; \bar t c \;+\; \bar t u)
 \; \; = \; \; 0.15 \; \; \; {\rm pb}. \label{sig4} 
\end{eqnarray}
Then, with the total luminosity 
of~${\cal L}_{int} = 70$~pb$^{-1}$ we get the number of events with a single
top production as follows:
\begin{eqnarray}
\begin{array}{l l c l }
 & N_t  &=&  10.5, \\
 & N_h(W \to 2jet) &=& 7.1, \label{sig6} \\
 & N_l(W \to e^{\pm} \nu \, + \,  \mu^{\pm} \nu) &=& 2.3. 
\end{array} 
\end{eqnarray}
Here $N_h$ and $N_l$ stand for the number of events with top quark decays via
pure hadronic or leptonic ($e + \mu$)  channels, respectively.

These and all forthcoming estimates are evaluated under assumption of 100\%
efficiency of registration of hadronic jets or leptons. We ignore also the
possible contributions from the background events.

Now we consider the upper limits on the anomalous constants
$\kappa_{\gamma}$ and $\kappa_z$, which one can evaluate from LEP-2 data. To
do this would require, in particular, the maximum negative value of the
contribution into cross section~(\ref{sig0}) resulted from the interference
term. As is seen from~(\ref{sig0}) such an requirement occurs at the following
constraints on the relative constants of $g_i$ and $z_i$:
\begin{eqnarray}
 g_1 z_2 = g_2 z_1 < 0. \label{const3}
\end{eqnarray}
It follows herefrom that $g_1 z_1 + g_2 z_2 = 1$. As a result the cross
section~(\ref{sig0}) becomes the function depending on two parameters, namely
$\kappa_{\gamma}$ and $\kappa_z$. Note, however, that the variation
of the magnitude of the interference term gives only small correction
to the forthcoming results on the constraints. It is explained by the small
magnitude of this term as compared to contribution due to photon and $Z$ boson
exchange.

Bearing in mind the different notations and normalizations used in the
literature, we present the evaluated constraints on anomalous constants in
terms of constraints on the corresponding branching ratios for the decays of
$t \to c(u) \gamma$ and $t \to c(u) Z$. We perform our analysis for the energy
of $\sqrt{s} = 184$~GeV as well as for other possible values of the total
energy of $e^+ e^-$ annihilation at LEP-2 collider and for corresponding total
luminosities as given below:
\begin{eqnarray}
\begin{array}{lccccl}
\sqrt{s} &=& 184 \;\; {\rm GeV} \;\; & {\cal L}_{e^+ e^-} &=& 70 \; \; 
 {\rm pb}^{-1}, \nonumber \\
\sqrt{s} &=& 192 \;\; {\rm GeV} \;\; & {\cal L}_{e^+ e^-} &=& 200 \; \; 
 {\rm pb}^{-1}, \label{lumin} \\ 
\sqrt{s} &=& 200 \;\; {\rm GeV} \;\; & {\cal L}_{e^+ e^-} &=& 100 \; \; 
 {\rm pb}^{-1}. \label{lum1} 
\end{array} 
\end{eqnarray}

Fig. 3 presents the evaluated constraints on the branching ratios of 
Br$(t \to c(u) \gamma)$ and Br$(t \to c(u) Z)$ (at 95~\% confidence level). 
We also take into account the
possibility to combine the statistics from all four experiments 
(ALEPH, DELPHI, L3, and OPAL) at the LEP-2 collider. The dashed curves in this
figure correspond to these constraints.

Note, that the contribution due to annihilation via photon is very small 
(see~Fig.~2). As a result at the energy of~$\sqrt{s} = 184$~çÜ÷ and the total
luminosity of ${\cal L}_{e^+ e^-} = 70$~pb$^{-1}$ it is impossible to improve 
the estimate~(\ref{dec1}), evaluated by the CDF collaboration. At the same time
one may expect two times improvement of the constraint on the branching ratio 
of $t$ quark decay into $Z$ boson:
\begin{eqnarray}
\sqrt{s} = 184 \;\;{\rm GeV} \; \Rightarrow 
\left\{ \begin{array}{lccl}
{\rm Br}(t \to (c+u) \; \gamma) &\leq & 3.2 \% & (95\% \; C.L.), \\ 
{\rm Br}(t \to (c+u) \; Z) &\leq & 18 \% & (95\% \; C.L.). 
\end{array} \right. \label{dec184}
\end{eqnarray}

The increasing of the total energy of $e^+ e^-$ annihilation will provide a
substantial improvement of the modern constraints~(\ref{dec1}) 
and~(\ref{dec2}) on the corresponding branching ratios of top quark decay both
in $Z$ boson and a photon. We present below the results for the energy of
$\sqrt{s} = 192 \div 200$~GeV, estimated for joint statistics from all four
experiments:
\begin{eqnarray}
\sqrt{s} = 192(200) \;{\rm GeV} \; \Rightarrow 
\left\{ \begin{array}{lccl}
{\rm Br}(t \to (c+u) \; \gamma) &\leq & 0.3 \% & (95\% \; C.L.), \\ 
{\rm Br}(t \to (c+u) \; Z) &\leq & 1 \% & (95\% \; C.L.). 
\end{array} \right. \label{dec200}
\end{eqnarray}

Note, that because of the different expected total luminosities
(see~(\ref{lum1})) one should expect much the same constraints on the
anomalous constants (the branching ratios) for both energies of
$\sqrt{s} = 192$~Gev and 200~GeV.

As it follows from our analysis even at current (1997~y.) run of 
$e^+ e^-$ collider LEP-2 it is possible to improve the modern constraints on
the parameters of anomalous FCNC interaction of the top quark.

It worth to note that to obtain the constraints like~(\ref{dec200}) from the
data of future run of FNAL collider one needs a rather high luminosity 
about ${\cal L}_{FNAL} \geq 1 \div 10$~fb$^{-1}$ (see, for 
example~\cite{th2,fut}).

\section {\bf Differential distributions }

As it mentioned above, the top-quark is produced very close to its threshold
production at the energies of LEP-2 collider (i.e. at~$\sqrt{s} \leq 200$~GeV).
This fact leads to practically fixed values of energies of the final 
$t$, $c(u)$, $b$ quarks and $W$ boson in the reaction~(\ref{etc2}):
\begin{eqnarray}
\left. \begin{array}{lcccl}
E_t & \simeq & \frac{s + m^2_t - m^2_c}{2 \sqrt{s}} & \simeq & m_t, \\
E_{c(u)} & \simeq & \frac{s - m^2_t + m^2_c}{2 \sqrt{s}} & \simeq & 
 \sqrt{s} - m_t, \\
E_b & \simeq & \frac{m^2_t - m^2_W + m^2_b}{2 m_t},  & & \\
E_W & \simeq & \frac{m^2_t + m^2_W - m^2_b}{2 m_t}. & & 
 \end{array} \right. \label{dif1} 
\end{eqnarray}

The corresponding differential distributions as the functions of the energies
of final particles in the reaction~(\ref{etc2}) are shown in Fig.~4. Note, the
considered single top production leads to rather specific topology of events.
This topology differs radically from that of possible background process of
$W^+ W^-$ pair production:
\begin{eqnarray}
e^+ \; e^- \; \to \; W^+ W^- \; \to \; 4jet. \label{bg1} 
\end{eqnarray}
Two jets from the reaction~(\ref{etc2}) have practically fixed values of the
energies. For example, at $\sqrt{s} =$~184~GeV one has: 
\begin{eqnarray*}
E_b \sim 70 \; \; {\rm GeV} \quad {\rm and} \quad
E_c \sim 10 \; \; {\rm GeV}.
\end{eqnarray*}
This specific behavior of the energy distributions of the jets, originating
from charmed and beauty quarks, are distinctly different from those
distributions from the background process~(\ref{bg1}).

We emphasize once again that the energy distributions of jets is actually
determined by the kinematics of $t \bar c (\bar u)$ pair production process
and be very weakly dependent on the parameters of the model of FCNC
interaction of top quark.

On the other hand, the angular distributions of the final particles in the
reaction~(\ref{etc2}) essentially depend on the model parameters. It can easy
seen from the equation for the differential cross section of 
$d \sigma / d \cos \vartheta$ for $t$ and $\bar c$ production in the
reaction~(\ref{etc1}): 
\begin{eqnarray}
\frac{d \sigma(e^+ e^- \to t \bar c)}{d \cos \vartheta} = 
\frac{ 3 \pi \alpha^2}{8 \; s} 
\Biggl ( 1 - \frac{m^2_t}{s} \Biggr )^2 
\biggl [ \chi_{\gamma} \; + \; \chi_z \; + \; \chi_{int} \biggr ], \label{dif2}
\end{eqnarray}
where $\vartheta$ is the outgoing angle of $t$ quark with respect to initial
electron in c.m.s. reference frame. The terms describing the annihilation via
photon~($\chi_{\gamma}$), $Z$--boson~($\chi_z$), and their 
interference~($\chi_{int}$) have the form as follows:
\begin{eqnarray}
\chi_{\gamma} & = & 2 \frac{m^2_t}{\Lambda^2} \kappa^2_{\gamma} e_q^2
\frac{s}{m^2_t} \biggl ( 1 + \frac{m^2_t}{s} \biggr ) 
(1 - \lambda \cos^2 \vartheta), \label{dif3} \\
\chi_z & = & \frac{\kappa^2_z}{2 \sin^4 2\vartheta_W \; (1-\frac{M^2_Z}{s})^2} 
 \nonumber \\
& \times & \Bigl [ (1+a^2_w)(1+\frac{m^2_t}{s}) 
(1 + \lambda \cos^2 \vartheta) - 4 a_w (z_1^2 - z_2^2) \cos \vartheta 
\Bigr ], \label{dif4} \\
\chi_{int} &=& 4 e_q \kappa_{\gamma} \kappa_z \Bigl ( \frac{m_t}{\Lambda} 
\Bigr  ) 
\frac{ a_w (g_1 z_1 + g_2 z_2) - 
 (g_1 z_1 - g_2 z_2) \cos \vartheta}
{\sin^2 2\vartheta_W \; (1-\frac{M^2_Z}{s})}, \label{dif5}
\end{eqnarray}
where $\lambda = (1 - m^2_t / s) / (1 + m^2_t / s)$.

Note, that at the energies of LEP-2 collider one has $\lambda \ll 1$.
Therefore, one can deduce immediately from the above-presented expression for
$d\sigma / d \cos \vartheta$ that the contributions into annihilation via 
photon or $Z$ boson weakly depend on $\cos \vartheta$ (see.~(\ref{dif3}) 
and~(\ref{dif4})). By contrast, the angular dependence of the interference
term ($ \sim (g_1 z_1 - g_2 z_2) \cos \vartheta$) is essentially dictated by a
choice of model parameters. For example, at $g_1 z_1 = -g_2 z_2$ this
dependence has a maximum character, while at $g_1 z_1 = + g_2 z_2$ this
contribution is independent of $\cos \vartheta$ at all (see.~(\ref{dif5})).
This fact can be used for evaluation more detailed constraints on the
parameters of the anomalous FCNC interaction of $t$ quark.

\section {\bf Conclusion }

In the present article we examine the possibility of investigation of the
anomalous $t$ quark interaction via flavor-changing neutral currents at the
energies of $e^+ e^-$ collider LEP-2.

We analyze the events with production of single $t$ quark. We show that 
one can improve about two times the modern constraints on the parameters of
anomalous FCNC interaction of top quark by use of the results of the 
current~(1997) run of LEP-2 collider at the energy of $\sqrt{s} = 184$~GeV 
and the total luminosity of~${\cal L} \simeq 70$~pb$^{-1}$. 
With increasing both of the total annihilation energy up to
$\sqrt{s} = 192 \div 200$~GeV and the total luminosity up to
${\cal L} = 100 \div 200$~pb$^{-1}$ one may expect to derive the constraints on
anomalous constants comparable to those which would be resulted from a future
run of FNAL collider.

We show that the final state particles in the reaction
$e^+ e^- \to \bar c t \to 4jet$ has the specific kinematics. Two final jets 
($c$--jet and $b$--jet) have practically fixed energies.
This topology differs radically from that of possible background process with
four jets production. Moreover, this kinematics are practically independent on
the parameters of the model. At the same time the angular distributions of the
final particles in the considered reaction of a single top production have a
noticeable dependence on the anomalous constants. This fact can be used for
the evaluation of additional constraints on the model parameters.

\section*{\bf Acknowledgements}

\noindent We are grateful to B.A.~Arbuzov, A.G.~Miagkov and M.M.~Shapkin
for useful discussions. This work was supported, in part, 
by Russian Foundation for Basic Research,
projects no.~96-15-96575.

\newpage

\begin{figure}[t]
\begin{center}
\epsfig{file=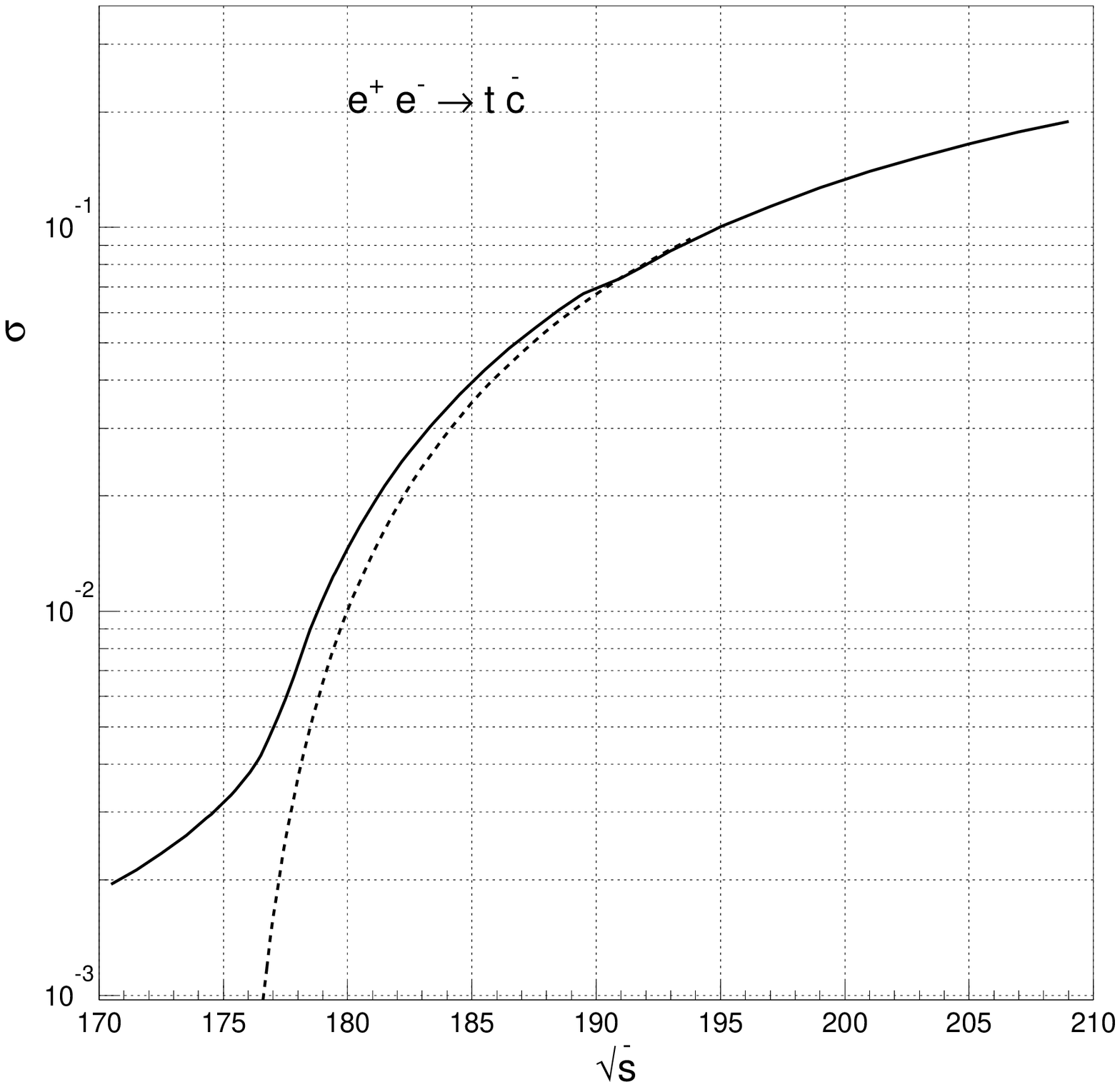,width=12cm,clip=}
 
\ccaption{}{
The cross sections for the process of $e^+ e^-$ annihilation in the
$t \bar c (\bar u$ pair in the reaction~(\ref{etc1}) (the dashed curve) 
and in the process~(\ref{etc2}) (the solid curve). The magnitudes of the
anomalous constants from~(\ref{const1}) and~(\ref{const2}) are used. The cross
section is in pb, while $\sqrt{s}$ in in GeV. 
}

\end{center}
\end{figure}

\newpage

\begin{figure}[t]
\begin{center}
\epsfig{file=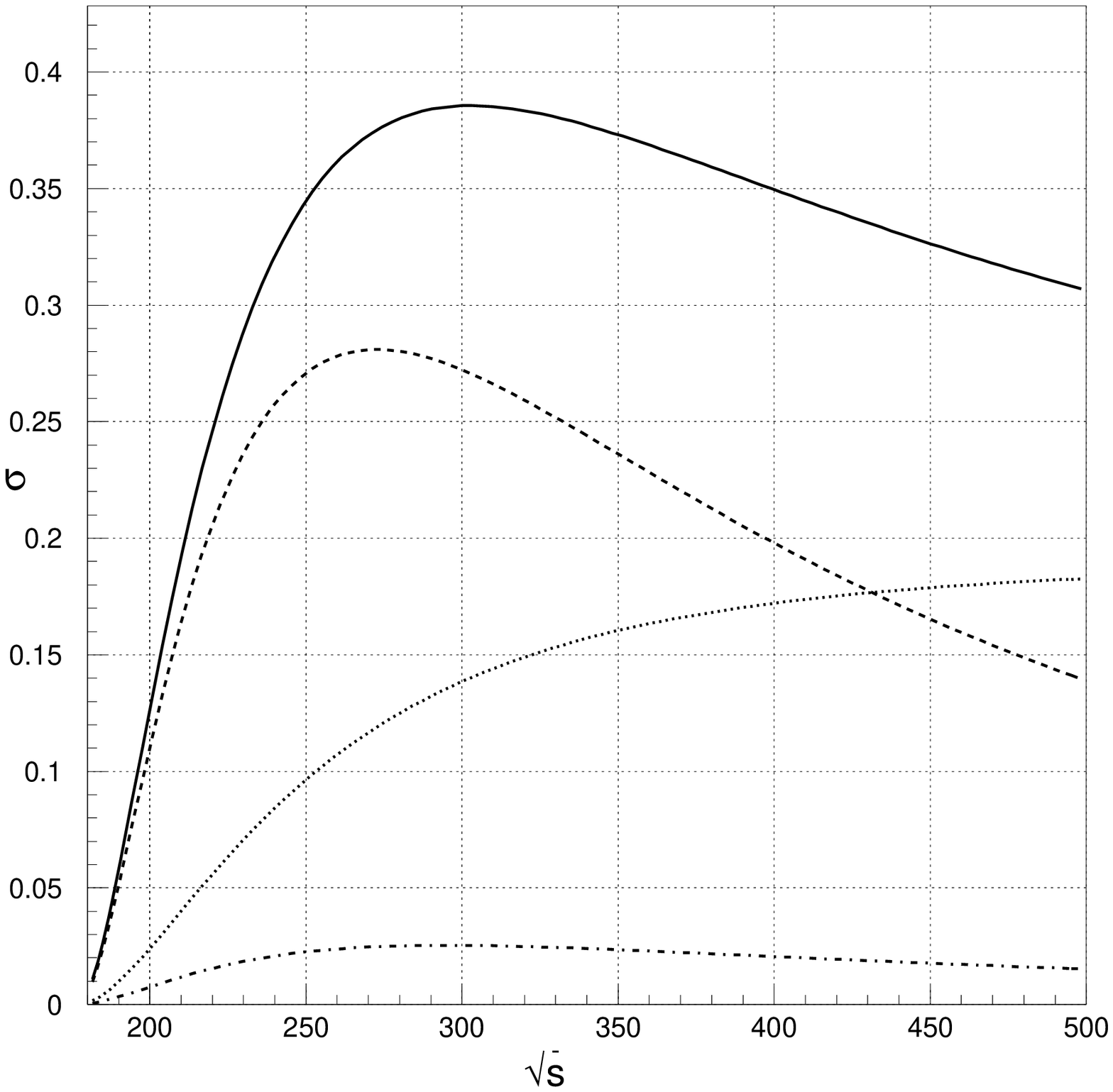,width=12cm,clip=}
 
\ccaption{}{
The cross sections for the reaction of $e^+ e^- \to t \bar c$  as a function
of the total energy of $\sqrt{s}$ (the solid curve). The dashed, dotted and 
dash-dotted curves correspond to the contributions from annihilation via a 
photon, $Z$--boson, and their interference, respectively. The cross
section is in pb, while $\sqrt{s}$ in in GeV. 
}
\end{center}
\end{figure}

\newpage

\begin{figure}[t]
\begin{center}
\epsfig{file=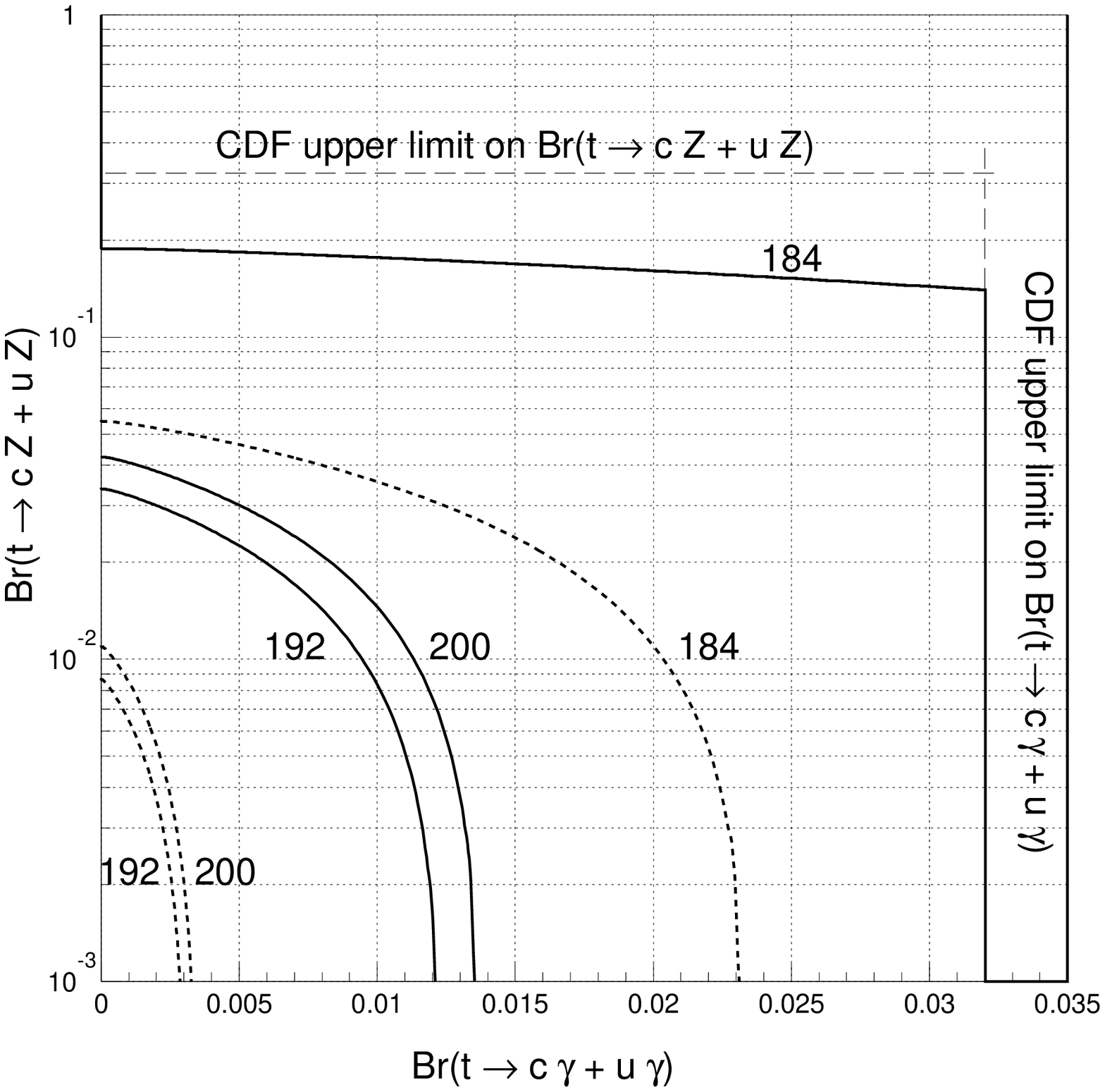,width=12cm,clip=}
 
\ccaption{}{
The upper limits (at 95 \% confidence level) on the branching ratios for the
decays $t \to (c+u) Z$ and $t \to (c+u) \gamma$ for the different values of
the total energy and luminosity of $e^+ e^-$ annihilation 
($\sqrt{s} = 184$~GeV and ${\cal L} = 70$~pb${}^{-1}$,
$\sqrt{s} = 192$~GeV and ${\cal L} = 200$~pb${}^{-1}$, and
$\sqrt{s} = 200$~GeV and ${\cal L} = 100$~pb${}^{-1}$). The dashed curves are
calculated with assumption of joint statistics from all four LEP--2 
experiments (i.e. ${\cal L}(184) = 280$~pb${}^{-1}$,
${\cal L}(192) = 800$~pb${}^{-1}$, and
${\cal L}(200) = 400$~pb${}^{-1}$).
}

\end{center}
\end{figure}

\newpage

\begin{figure}[t]
\begin{center}
\epsfig{file=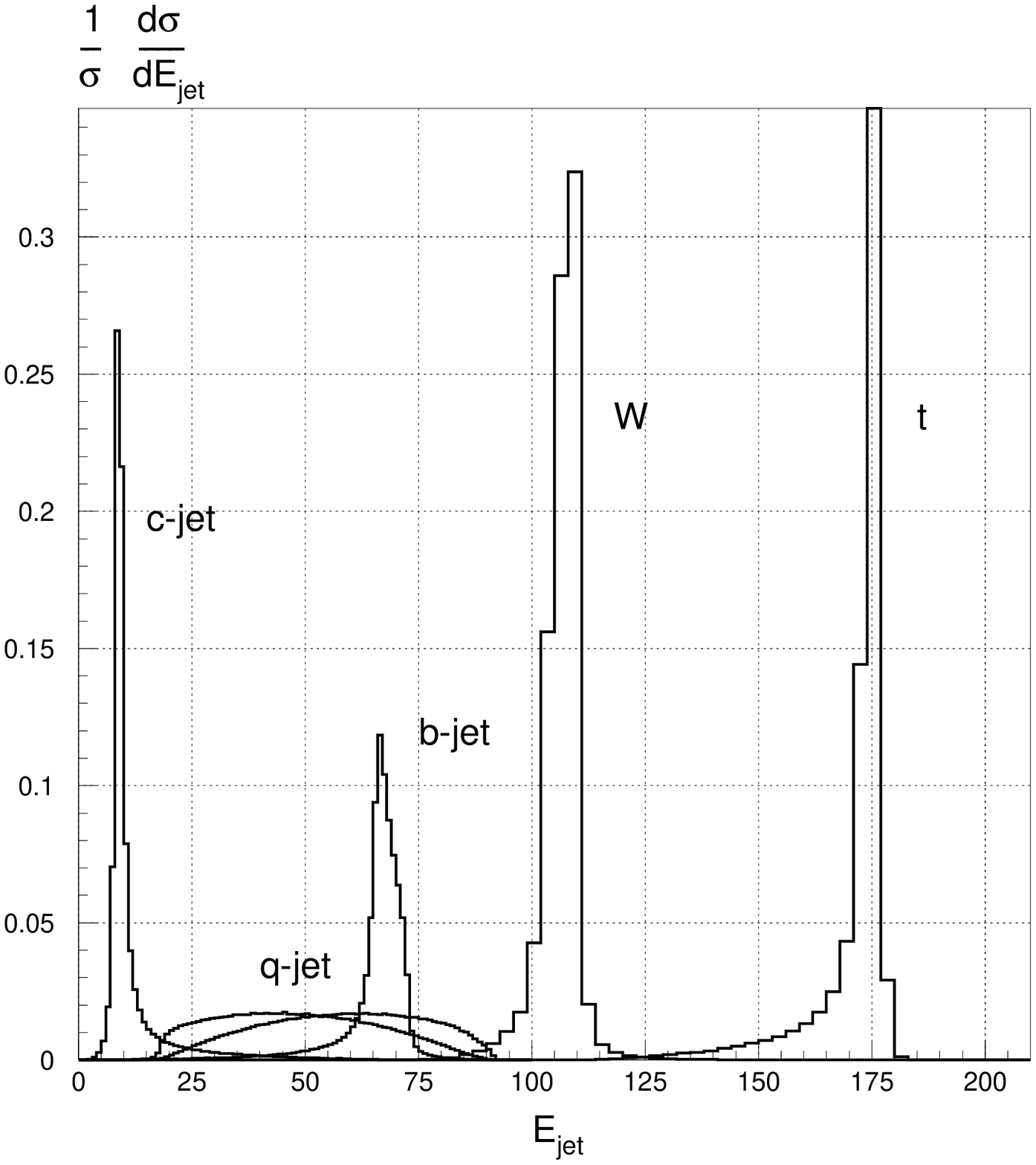,width=12cm,clip=}
 
\ccaption{}{
Distribution on energies of the final state particles from the 
reaction~(\ref{etc2}). The curves with labels "q--jet" corresponds to jets
from $W$ boson decays. The energy of jets ($E_{jet}$ is in GeV, while the 
cross section $(1/\sigma) d \sigma / E_{jet}$ is in~GeV${}^{-1}$.
}

\end{center}
\end{figure}

\end{document}